\documentclass{aa}  
\usepackage{natbib}
\bibpunct{(}{)}{;}{a}{}{,}
\usepackage{graphicx}
\usepackage{float}
\usepackage{color}
\usepackage{hyperref}
\usepackage{siunitx}
\usepackage{comment}
\usepackage{xspace}
\usepackage{multirow}

\usepackage{txfonts}
\usepackage{hyperref}
\makeatletter
\renewcommand*\aa@pageof{, page \thepage{} of \pageref*{LastPage}}
\makeatother

\DeclareMathOperator\arcsinh{arcsinh}
\newcommand{\atrous}{{\it à trous}\xspace}

\usepackage[normalem]{ulem}

\begin{document} 

   \title{Image enhancement with wavelet-optimized whitening}
   \author{F. Auchère\inst{\ref{inst:ias}}
        \and
        E. Soubrié\inst{\ref{inst:ias}}
        \and
        G. Pelouze\inst{\ref{inst:ias}}
        \and
        É. Buchlin\inst{\ref{inst:ias}}
          }
        \institute{Université Paris-Saclay, CNRS, Institut d'Astrophysique Spatiale, 91405, Orsay, France\label{inst:ias}}

   \date{Received 01 November 2022; accepted 02 December 2022}

 
  \abstract
   {Due to its physical nature, the solar corona exhibits large spatial variations of intensity that make it difficult to  simultaneously visualize the features present at all levels and scales. Many general-purpose and specialized filters have been proposed to enhance coronal images. However, most of them require the ad hoc tweaking of parameters to produce subjectively good results.}
   {Our aim was to develop a general purpose image enhancement technique that would produce equally good results, but based on an objective criterion.}
   {The underlying principle of the method is the equalization, or whitening, of power in the \atrous wavelet spectrum of the input image at all scales and locations. An edge-avoiding modification of the \atrous transform that uses bilateral weighting by the local variance in the wavelet planes is used to suppress the undesirable halos otherwise produced by discontinuities in the data.}
   {Results are presented for a variety of extreme ultraviolet (EUV) and white light images of the solar corona. The proposed filter produces sharp and contrasted output, without requiring the manual adjustment of parameters. Furthermore, the built-in denoising scheme prevents the explosion of high-frequency noise typical of other enhancement methods, without smoothing statistically significant small-scale features. The standard version of the algorithm is about two times faster than the widely used   multiscale Gaussian normalization (MGN). The bilateral version is slower, but provides significantly better results in the presence of spikes or edges. Comparisons with other methods suggest that the whitening principle may correspond to the subjective criterion of most users when adjusting free parameters.}
{}
\maketitle

\section{Introduction\label{sec:introduction}}

\begin{figure*}
    \centering
    \includegraphics[width=\textwidth]{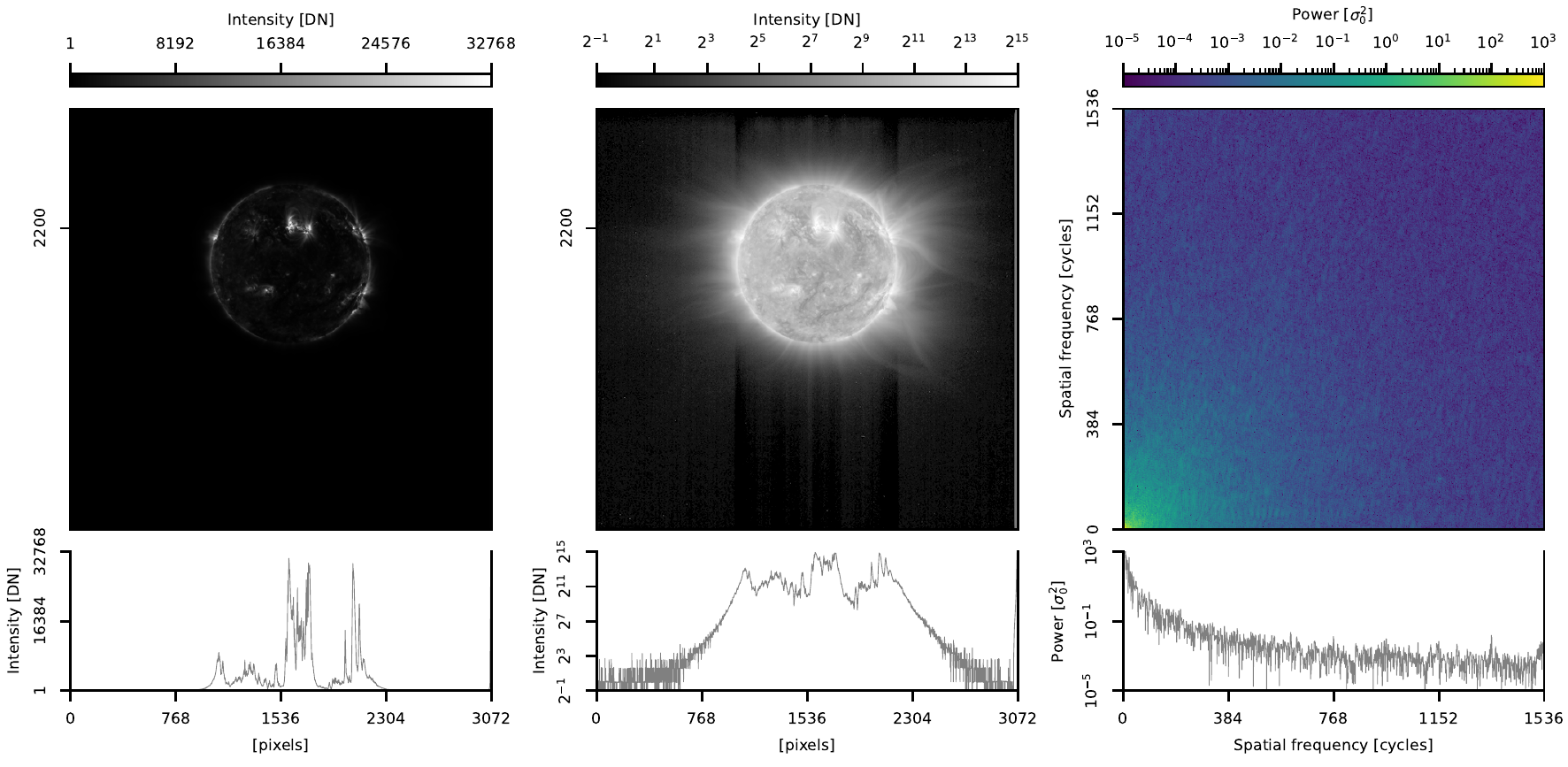}
    \caption{Dynamic range and spatial scales in a solar EUV image. Left: Fifteen-bit image taken by the Full Sun Imager (FSI) channel of EUI at \SI{17.4}{\nano\metre} on 2022~March~17 at 06:00:45~UT, displayed in linear scale. Only the brightest features are visible. The profile is taken along the row marked by the side ticks. Middle: Same image after logarithmic scaling. Faint off-limb details are revealed, but the local contrast is diminished. Right: Fourier power spectrum of the variance normalized image. The profile is that of the bottom row. The larger the spatial scale, the more power (and thus variability) in the image.}
    \label{fig:stretch_fourier}
\end{figure*}

Modern image sensors, like those used in the Extreme Ultraviolet Imager ~\citep[EUI,][]{Rochus2020} on board the Solar Orbiter mission~\citep{Muller2020}, have dynamic ranges (ratio of the digital range to the read noise) of $2^{12}$ or more. As long as the observed scenes use the entire detector range, the typical $2^8$ dynamic of monitors is insufficient to display the recorded images unmodified without losing information (left panel of Fig.~\ref{fig:stretch_fourier}). Therefore, most of the time the intensity values are  rescaled to 8 bits prior to display using a  nonlinear function (or stretch), usually a power $1/\gamma<1$ ($\gamma$-stretch) or a logarithm of the intensity (middle panel of Fig.~\ref{fig:stretch_fourier}). It is worth noting that a square root transformation ($\gamma=2$) is sometimes applied on board (e.g., on EUI) for it provides optimum intensity sampling for images dominated by Poisson statistics~\citep{Nicula2005}. Nonlinear intensity transformations alone,  however, are  not sufficient to bring out all the information present in the data. Filtering and enhancement is a crucial step in the analysis of scientific images for it can reveal features and phenomena that would otherwise remain undetected.

As illustrated in the right column of Fig.~\ref{fig:stretch_fourier}, the Fourier power spectrum of a typical solar EUV image follows a power-law-like distribution, with more power at low frequencies and less power at high frequencies. As per Parseval's theorem, the mean power of the Fourier spectrum is equal to the variance of the signal. Thus, the larger (respectively lower) Fourier power at lower (respectively higher) frequencies correspond to larger (respectively lower) signal variance at larger (respectively smaller) spatial scales. The middle column of Fig.~\ref{fig:stretch_fourier}, which is remapped using a logarithm intensity transformation, indeed shows that the largest intensity variations come from the global scales. However, while bringing out the faint  large-scale features, this simple intensity remapping produces a washed out image with poor local contrast. Intensity variations in the active regions are mapped to a much smaller range of output values than with the linear scaling of the left column. This is why many enhancement methods are designed to reduce the signal variance at low frequencies and to increase it at high frequencies.

Specialized background subtraction methods~\citep[e.g.,][]{Morgan2006, Patel2022} can be used to this end, but they are typically limited to the off-disk emission. One of the earliest general-purpose methods, the unsharp mask, is a simple high-pass filter that was developed to improve the quality of photographic prints by analog composition of a negative with a blurred positive~\citep{Yule1944, Schreiber1970, Levi1974}. Numerically, the convolution kernel is generally taken to be Gaussian, but the frequency response of the filter can be tailored to arbitrary shapes in the Fourier space. However, Fourier filtering techniques are  limited in that they ignore the local variations of the image properties, unless the input is segmented into independent tiles.

On the contrary, multiscale approaches are able to modify the spectral content differently at different positions in the input image. The detail layer of the multiscale Gaussian normalization~\citep[MGN,][]{Morgan2014} is essentially a weighted sum of unsharp-masked images of increasing Gaussian kernel widths, normalized to their local standard deviations and remapped to the [-1, 1] range. MGN is successful in equalizing the variances over a range of scales, but also enhances the high-frequency noise in doing so (see bottom right panel of Fig.~\ref{fig:wow_nafe_mgn}), a common issue with image sharpening techniques.

The decomposition of images into scales using a wavelet transform offers the possibility to simultaneously denoise~\citep{Starck1994, Murtagh1995} and modulate the output power spectrum. The wavelet packet equalization scheme of \cite{Stenborg2003, Stenborg2008} consists in a two-level \atrous wavelet decomposition, followed by local noise reduction and weighted synthesis.

Enhancement methods typically include free parameters that need to be adjusted by the user: the kernel width for the unsharp mask; the kernel widths, variance regularization function, and synthesis weights in MGN; and  the noise threshold and synthesis weights in \cite{Stenborg2003}. This provides flexibility in tuning the output, with the drawback of subjectivity in deciding what good results are. 

In the present article we propose a transform that whitens the power spectrum, resulting in equal variance at all spatial frequencies (section \ref{sec:method}). While not the only possible choice of normalization, a comparison of the results (section \ref{sec:results}) with other methods suggests that images considered to be  well enhanced tend to have a white spectrum. We summarize our work in section~\ref{sec:conclusions}.

\section{Method\label{sec:method}}

\subsection{À trous transform\label{sec:transformg}}

\begin{figure*}
    \centering
    \includegraphics[width=\textwidth]{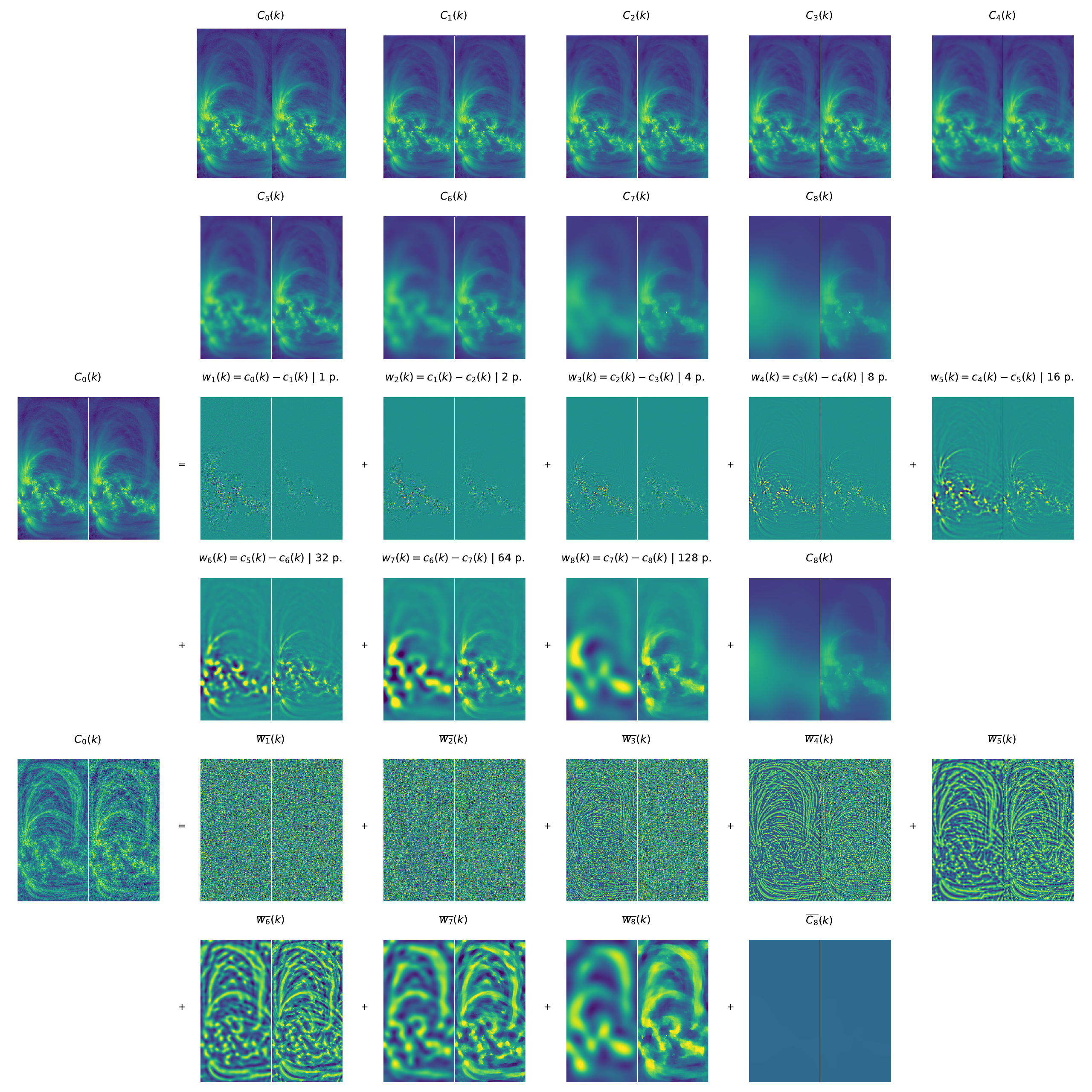}
    \caption{{\it A trous} wavelet transform and coefficients whitening. Two top rows: Chained convolutions $c_s(k)$ of the input image $c_0(k)$ with \atrous kernels dilated by a factor of  two at each step. The left half of each panel corresponds to regular convolutions; the right half to bilateral filtering. Two middle rows: Corresponding wavelet coefficients $w_{s+1}(k) = c_s(k)-c_{s+1}(k)$, that is the difference between successive convolutions or bilateral filterings. The last coefficient is the result of the last convolution or bilateral filtering. The original image ($c_0(k)$, left) is the sum of all coefficients. The bilateral transform preserves the edges in the coarse images (top rows). Two bottom rows: Whitened coefficients, {i.e.}, normalized to the local power. The filtered image (left) is the sum of all whitened coefficients. For visualization purposes, the convolutions and the coefficients are displayed using two different color scales.}
    \label{fig:wavelets-wow-standard}
\end{figure*}

We summarize below only those aspects of the \atrous transform necessary to understand the proposed method. We refer the reader to \cite{Holschneider1989, Shensa1992, Starck2002} and references therein for material on wavelet theory and the \atrous transform in particular. See also \cite{Stenborg2003} for a concise introduction and application to images of the solar corona.

The wavelet transform of a discrete signal $f(l)$ is defined by
\begin{equation}
    w_s(k)=\frac{1}{\sqrt{s}}\sum_l\psi^*\left(\frac{l-k}{s}\right)f(l)
,\end{equation}
where $s$ is a scaling factor. In the case of the \atrous transform, the real-valued wavelet $\psi$ satisfies the relationship
\begin{equation}
    \frac12\psi\left(\frac{x}2\right)=\phi(x)-\frac12\phi\left(\frac{x}2\right)
\end{equation}
with $\phi(x)$ a scaling function that obeys the dilation equation
\begin{equation}
    \frac12\phi\left(\frac{x}{2}\right)=\sum_l h(l)\,\phi(x-l)\label{eq:dilation}
,\end{equation}
where $h$ is a low-pass filter kernel. We can then demonstrate ~\citep{Shensa1992} that the \atrous wavelet transform can be computed by the following iteration, initialized with the input image $I(k)=c_0(k)$:
\begin{align}
    c_{s+1}(k) & = \sum_{l\in\Omega}c_s(k+2^s l)\,h(l) = h_s(k) * c_s(k)\label{eq:convolution},\\
    w_{s+1}(k) & = c_s(k) - c_{s+1}(k).
\end{align}
Here $\Omega$ denotes the support of the compact kernel $h$ and $*$ the convolution product. The set $\{w_1(k)\ldots w_S(k), c_S(k)\}$ forms the wavelet transform of the input image, with $S$ the largest scale compatible with the data. At scale $s$, the data values weighted by $h(l)$ are separated by $2^s$, which is equivalent to a convolution with a sparse (with holes, \atrous in French) kernel $h_s(k)$. We note that this iteration lends itself naturally to an implementation in which successive scales are recursively computed by convolutions of the compact kernel with interleaved subsamplings of one point out of two on each axis.

The successive convolutions form increasingly smooth representations $c_s(k)$ of the data (left half images in the first two rows of Fig.~\ref{fig:wavelets-wow-standard}). The input image $c_0(k)$ used in this example was acquired with the \SI{17.4}{\nano\metre} high-resolution EUV channel of EUI (HRI$_\text{EUV}$, top left panel of Fig.~\ref{fig:wow_stages}) on 2022 March 17 at 04:02:57~UT, when Solar Orbiter was at 0.38~AU from the Sun. It is the one closest in time to the FSI image of Fig.~\ref{fig:stretch_fourier}. Its wavelet transform with $S=8$ is represented in the left half images of the third and fourth rows. The input image can be synthesized with
\begin{equation}
c_0(k) = c_S(k) + \sum_{s=1}^S w_s(k)
\label{eq:synthesis}
.\end{equation}
In the remainder of this paper we use a basic cubic spline for the scaling function $\phi$, which corresponds (Equation~\ref{eq:dilation}) to the one-dimensional kernel $h=\left[1, 4, 6, 4, 1\right]/16$, its n-dimensional counterparts being obtained by iterating $h^{T}h$. Other choices are possible, but the $B_3$-spline is smooth, is continuously differentiable, and  has minimum curvature and a compact support, which makes it both robust to oscillations and computationally efficient~\citep[see, e.g.,][]{Unser1999}. We use mirror boundary conditions so that $c(k+N)=c(N-1-k)$ with $N$ the number of pixels on a given axis.

The wavelet coefficients (third and fourth row of Fig.~\ref{fig:wavelets-wow-standard}) can be manipulated to achieve a variety of results after synthesis. Denoising will be treated in section~\ref{sec:denoising}. In the next section, we   discuss multiscale contrast enhancement.

\subsection{Whitening\label{sec:whitening}}

\begin{figure*}
    \centering
    \includegraphics[width=\textwidth]{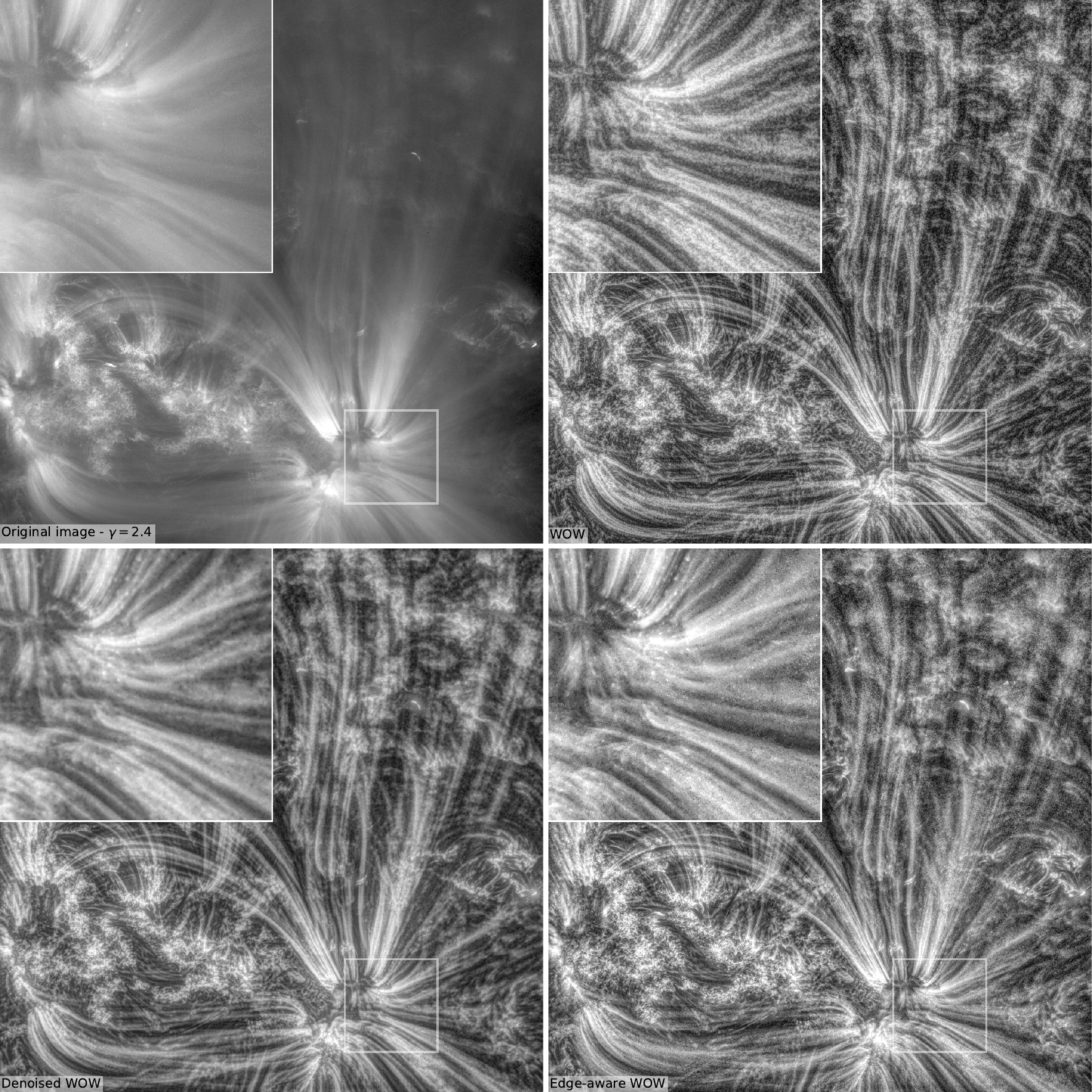}
    \caption{Three levels of WOW processing. Top left: Original $\gamma$-stretched ($\gamma=2.4$) HRI$_\text{EUV}$ image. The inset (top left quarter) shows an enlargement of the area within the white square. Top right: Output of the WOW algorithm. Bottom right: Same with added denoising. Bottom right: Output of the edge-aware denoised version of the algorithm. For all images the gray scale maps to the central 99.8\% of values. An animated version of this figure is available online.}
    \label{fig:wow_stages}
\end{figure*}

It is possible to enhance features at a given scale, or more generally to modify the relative importance of different scales, by multiplying the corresponding coefficients $c_k^s$ with ad hoc weights $\beta_s$ at the image synthesis stage, so that Equation~\ref{eq:synthesis} becomes
\begin{equation}
\overline{c_0}(k) = c_S(k) + \sum_{s=1}^S \beta_s \, w_s(k)
\label{eq:weighted_synthesis}
.\end{equation}
It is possible, for example, to suppress large-scale gradients by setting the last weight to zero. This approach is used in the wavelet packet equalization method of~\cite{Stenborg2003}. While it yields interesting results, it also  produces undesirable ringing around sharp and bright features (see, e.g., the right panel of their Fig. 4). Nonetheless, building on this idea, we wondered  whether the weights, instead of being arbitrary, could be dictated by the data content to produce an optimal output. Figure~\ref{fig:stretch_fourier} suggests that a possible scheme consists in equalizing the power at all scales (i.e.,  whitening the power spectrum). However, using one global weight per scale ignores the information contained in the spatial variations of power in the wavelet spectrum. In order to equalize the spectrum in both the spectral and spatial dimensions, we normalize the coefficients using an estimate of the local mean power $\overline{P_s}(k)$ obtained by convolution of the wavelet power by the \atrous kernel. The whitened coefficients $\overline{w_s}(k)$ are given by
\begin{equation}
\overline{w_s}(k)=\frac{w_s(k)}{\sqrt{\overline{P_s}(k)}},\quad\mathrm{with}\quad \overline{P_s}(k) = w_s(k)^2*h_s(k)
\label{eq:whitening}
.\end{equation}
Unweighted synthesis of the whitened coefficients results in the wavelet-optimized whitening (WOW) enhanced image:
\begin{equation}
\overline{c_0}(k) = \overline{c_S}(k) + \sum_{s=1}^S \overline{w_s}(k)
\label{eq:whitening_synthesis}
.\end{equation}

We note   the similarity between Equation~\ref{eq:whitening} and Equations~1 and 2 of~\cite{Morgan2014}: both represent convolutions of the original image by variable-sized kernels normalized to a quantity analogous to a local standard deviation. However, the numerator in Equation~1 of~\cite{Morgan2014} is an unsharp mask (i.e., the difference between the original image and its convolution by a Gaussian), while in Equation~\ref{eq:whitening} it is an \atrous wavelet coefficient (i.e.,  the difference between two successive chained convolutions by a scaling function). As we show   in section~\ref{sec:comparison}, this difference between the two methods results in significantly different output.

Figure~\ref{fig:wow_stages} shows a $\gamma$-stretched high-resolution \SI{17.4}{\nano\metre} EUI image (top left), and the corresponding output of the WOW process (top right). Only the brightest regions of the input image would be visible without $\gamma$-stretching, as in the left panel of Fig.~\ref{fig:stretch_fourier}.  Multiscale whitening tends to equalize the variance at all spatial frequencies and locations, which attenuates the large-scale variations and comparatively reinforces the small-scale structures. The large-scale contrast is reduced, while the small-scale contrast is increased. The enlargement of the region in the  white square shows that the noise, which behaves like a small-scale feature, is amplified in the process, especially in the originally faint regions where  the noise contributes the most to the signal variance. However, the \atrous wavelet transform offers a powerful framework for denoising.

\subsection{Denoising\label{sec:denoising}}

\begin{table}
\caption{Gain $g$ and read noise $r$ for the detectors used in this paper (Equation~\ref{eq:sigma_gaussian_poisson}).}
\begin{tabular}{lccl}
\hline\hline
Instrument & Gain & Read noise & Reference\\
& [DN/ph] &  [DN] & \\
\hline
FSI$_{304}$& 3.88 & 1.50 & \cite{Gissot2022}\\
HRI$_\text{EUV}$ & 5.28 & 1.50 & \cite{Gissot2022}\\
AIA$_{174}$& 1.12 & 1.15 & \cite{Boerner2012}\\
LASCO C2 & 0.07 & 0.3 & \cite{Brueckner1995}\\
\hline
\end{tabular}
\label{tab:detectors_characteristics}
\end{table}

The data can be denoised by attenuation of the wavelet coefficients that are not statistically significant, followed by synthesis~\citep{Starck1994, Murtagh1995}. This is achieved by comparing the $w_s(k)$ with the amplitudes expected in the presence of noise only. In the case of Gaussian white noise of unit standard deviation, the amplitudes at scale $s$ are Gaussian-distributed with standard deviation $\sigma_s^1$, which can easily be estimated numerically~\citep{Starck2002}. In the presence of noise of standard deviation $\sigma$, the significant coefficients are those for which
\begin{equation}
\left|w_s(k)\right|>n_s\,\sigma_s,\quad\mathrm{with}\quad \sigma_s=\sigma\,\sigma_s^1
\label{eq:coeff_signif}
,\end{equation}
where $n_s$ sets the chosen significance level at scale $s$. The coefficients can then be weighted by a function $\alpha(n_s\,\sigma_s, w_s(k))$ of their values relative to $n_s\,\sigma_s$. Hard thresholding corresponds to
\begin{equation}
    \alpha(n_s\,\sigma_s, w_s(k))=
    \begin{cases}
      1 & \text{if} \quad \left|w_s(k)\right|>n_s\,\sigma_s\\
      0 & \text{otherwise}
    \end{cases}
\label{eq:hard_thresholding},
\end{equation}
but tends to produce ringing in the synthesized image. There are several possibilities for smooth soft-thresholding functions. We chose to weight the coefficients by their probability of chance occurrence, so that
\begin{equation}
    \alpha(n_s\,\sigma_s, w_s(k))=\mathrm{erf}\left(\frac{\left|w_s(k)\right|}{n_s\,\sigma_s}\right).
\label{eq:soft_thresholding}
\end{equation}
In addition to the Gaussian component corresponding to the read and thermal noises within the sensor, the data also contains the Poisson-distributed component of the photon shot noise. The variance of the Poisson component is proportional to the intensity and is thus variable across the image. In the case of pure Poisson statistics, the transform
\begin{equation}
T[I(k)] = \sqrt{I(k) + \frac{3}{8}}
\end{equation}
results in data with Gaussian noise of unit variance~\citep{Anscombe1948} for numbers of counts greater than about ten. It was generalized by~\cite{Murtagh1995} for the presence of additional Gaussian noise in the input.

Following \cite{Starck1997, Stenborg2003}, the alternative approach adopted in this paper to account for the spatial dependence of the noise consists in replacing $\sigma_s$ in Equations~\eqref{eq:coeff_signif}--\eqref{eq:soft_thresholding} by a spatially variable $\sigma_s(k)=\sigma(k)\,\sigma_s^1$. The local standard deviation of the noise $\sigma(k)$ can be estimated from the first wavelet coefficients~\citep{Starck1997, Stenborg2003} or, preferably,  if the detector gain $g$ and read noise standard deviation $r$ are known, with
\begin{equation}
\sigma(k)=\sqrt{gI(k) + r^2}
\label{eq:sigma_gaussian_poisson}
.\end{equation}
In the end, for combined denoising and whitening, equation~\ref{eq:whitening} becomes
\begin{equation}
\overline{w_s}(k)=\frac{\alpha(n_s\,\sigma_s(k), w_s(k))\,w_s(k)}{\sqrt{\overline{P_s}(k)}}
\label{eq:denoised_whitening}
;\end{equation}
the expressions for $\overline{P_s(k)}$ and the synthesis (Equation~\ref{eq:whitening_synthesis}) remain unchanged.

The bottom left panel of Fig.~\ref{fig:wow_stages} shows the denoised whitening of the top left input image, to be compared with the regular whitening (Section~\ref{sec:whitening}) of the top right panel. The detector gain and read noise (Equation~\ref{eq:sigma_gaussian_poisson}) for HRI$_\text{EUV}$ are given in Table~\ref{tab:detectors_characteristics}. Noise being inherently high-frequency, it is usually significant only in the first few scales. In this example, we used $n_s=\{5, 2, 1\}$ for the first three scales and $n_s=0$ for the others. This choice is arbitrary; larger values yield smoother output at the expense of losing fainter details. 

\subsection{Edge-aware transform\label{sec:edge-aware}}

In the \atrous wavelet transform each convolution by the scaling function (Equation~\ref{eq:convolution}) produces halos around sharp peaks and edges. These halos are reinforced as genuine features by the whitening process, which tends to produce glows and/or gradient reversals. As proposed by~\cite{Hanika2011}, this unwanted side effect can be mitigated by replacing the convolution by bilateral filtering in Equation~\eqref{eq:convolution}. Range filtering is analogous to spatial filtering, but instead of being a function of spatial distance, the kernel weights are a function of the disparity of values. The bilateral filter~\citep{Tomasi1998} combines both spatial and range filtering, so that Equation~\eqref{eq:convolution} becomes
\begin{equation}
    c_{s+1}(k) = \frac{\sum_{l\in \Omega} c_s(k+2^sl)\,h(l)\,b(k, k+2^sl))}{\sum_{l\in \Omega}{h(l)\,b(k, k+2^sl)}}
,\end{equation}
where the denominator ensures unit normalization. A classical choice for the range filter $b$ is a Gaussian
\begin{equation}
    b(k, k+2^sl)=\exp{\left(-\frac12\left(\frac{c_s(k)-c_s(k+2^sl)}{\nu_s}\right)^2\right)}.
\end{equation}
Following the idea of equalization, we use for $\nu_s(k)^2$ the local variance at scale $s$, so that the range weights do not depend upon the absolute values of the wavelet coefficients. For computational efficiency, this is approximated by
\begin{equation}
\nu_s(k)^2=h_s(k)*c_s(k)^2 - \left(h_s(k)*c_s(k)\right)^2
.\end{equation}

The resulting edge-aware transform is illustrated in the right  halves of the panels of Fig.~\ref{fig:wavelets-wow-standard}. Compared to the regular transform, the edges are preserved in the coarse images (top two rows) by the bilateral filtering, and the power is correspondingly reduced in the coefficients (middle two rows). For denoising with the edge-aware transform, the corresponding standard deviations $\sigma_s^1$ of Gaussian white noise were estimated numerically. The bottom right panel of Fig.~\ref{fig:wow_stages} shows the output of the denoised edge-aware WOW algorithm for the top left input image. With the same set of significance levels $n_s$ as for the regular transform (bottom left panel), the output is sharper, but denoising is less pronounced. An animated version of Fig.~\ref{fig:wow_stages} containing 899 frames acquired on 2022 March~17 from 03:18:00 to 04:02:57~UT is available online.

\section{Results\label{sec:results}}

\subsection{Three examples}

\begin{figure*}
    \centering
    \includegraphics[height=0.9\textheight]{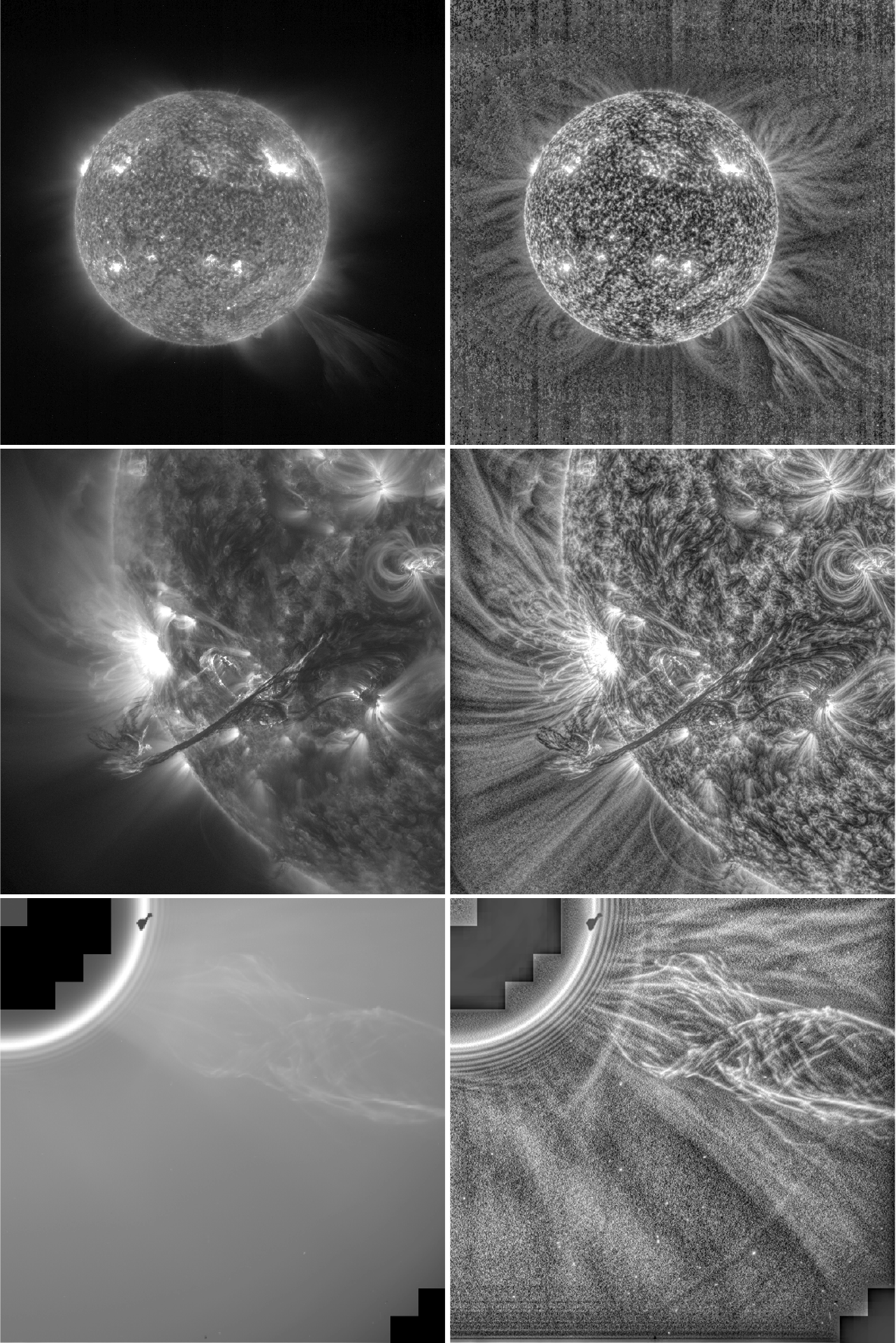}
    \caption{Examples of WOW processed images. For each row, the left image is the original $\gamma$-stretched with $\gamma=2.4$, the right image is the output of the edge-aware denoised version of the algorithm. Top: Crop of the \SI{30.4}{\nano\metre} FSI image taken on 2022~March~21 at 06:30:15~UT. Middle: \SI{17.1}{\nano\metre} AIA image taken on 2012 August 31 at 19:44:35~UT. Bottom: LASCO C2 image taken on 1998~June~2 at 13:31:06~UT.}
    \label{fig:wow_examples}
\end{figure*}

Figure~\ref{fig:wow_examples} shows three examples of images processed with the edge-aware version of the WOW algorithm. The original $\gamma$-stretched ($\gamma=2.4$) images are in the left column and the linearly stretched processed images,  all obtained using the same denoising significance levels $n_s=\{5, 3, 1\} $,  are in the right column. The gain and read noise for the different detectors are given in Table~\ref{tab:detectors_characteristics}. For all images, the black and white points of the linear grayscale are mapped respectively to the 0.1 and 99.9 percentiles. In the FSI \SI{30.4}{\nano\metre} image (top row) the rapid off-disk intensity fall-off masks the faint outer coronal structures (see also Fig.~\ref{fig:stretch_fourier}). In the top right image, the whitening of the wavelet spectrum enhances the local contrast and reveals the fainter extensions. The method also accentuates imperfections in the calibration (vertical bands and pattern at the top and bottom edges) because they are spatially correlated, and therefore are not considered as noise by the algorithm. The Atmospheric Imaging Assembly~\citep[AIA,][]{Lemen2012} image (middle row) can be compared to Fig.~2 of ~\cite{Druckmuller2013} processed with noise adaptive fuzzy equalization (NAFE, see next section). The two images are visually similar, with WOW producing as sharp and contrasted an output without the {ad hoc} adjustment of parameters nor the addition of a $\gamma$-stretched component. In images from white-light coronagraphs~\citep[like that in the bottom left of Fig.~\ref{fig:wow_examples}, from LASCO C2,][]{Brueckner1995}, the dust corona and stray light form a diffuse background that dominates the K-corona. By equalizing the power across spatial frequencies, this component is attenuated with respect to smaller-scale features resulting, in the bottom right image, in the enhancement of the twisted filamentary structure of the erupting prominence. Comparison with the output of the wavelet packets equalization method ~\citep[Fig.~4 of ][]{Stenborg2003} shows that WOW avoids the ringing around the bright and sharp features, and also enhances the disturbances in the surrounding K-corona. As already mentioned, the algorithm treats coherent features irrespective of their solar or instrumental nature. For optimal results, artifacts like the inner diffraction rings or the long curved streaks should be either masked or calibrated out~\citep{Morgan2015, Lamy2020, Lamy2022} prior to enhancement.

\subsection{Comparison with NAFE and MGN\label{sec:comparison}}

\begin{figure*}
    \centering
    \includegraphics[height=0.9\textheight]{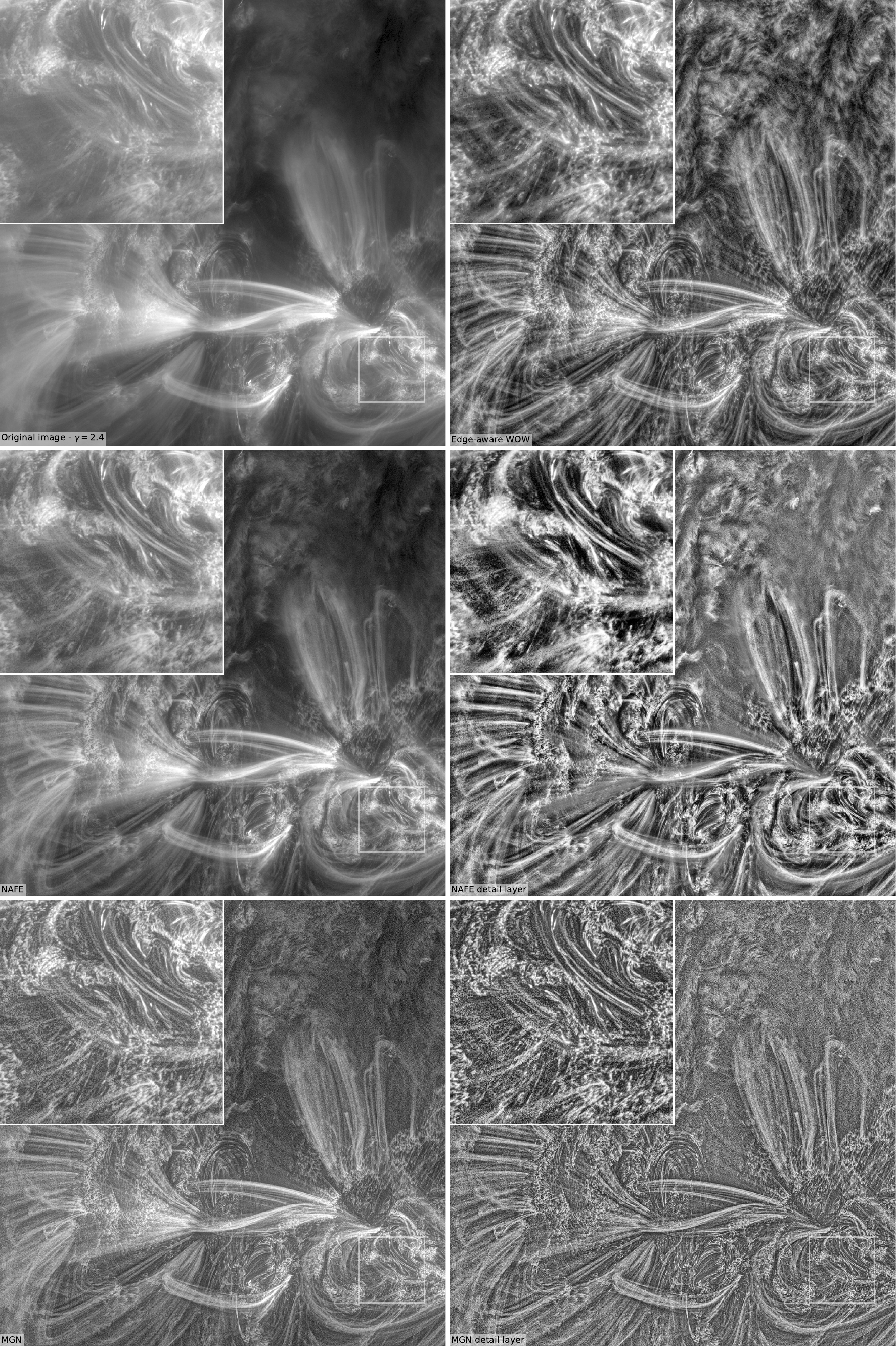}
    \caption{Comparison of the WOW, NAFE, and MGN algorithms. Top left: Original $\gamma$-stretched ($\gamma=2.4$) HRI$_\text{EUV}$ image. The inset (top left quarter) shows an enlargement of the area within the  white square. Top right: Output of the edge-aware denoised version of the algorithm. Middle row: NAFE (left) and corresponding detail layer (right). Bottom row: MGN (left), and corresponding detail layer (right). In NAFE and MGN, the $\gamma$-stretched original image (top left) is added to their respective detail layers (right column). The left column thus corresponds to: no enhancement, added NAFE detail layer, added MGN detail layer. The standard WOW output does not include an arbitrary $\gamma$-stretched component. An animated version of this figure is available online.}
    \label{fig:wow_nafe_mgn}
\end{figure*}

We chose to compare the output of WOW with that of NAFE~\citep{Druckmuller2013} and MGN~\citep{Morgan2014}, two algorithms commonly used in the solar physics community. The output $O(k)$ of both NAFE and MGN is the weighted sum of the $\gamma$-stretched input $I(k)$ and of the filtered image $F(k)$ proper:
\begin{equation}
O(k) = w \left(\frac{I(k) - a_0}{a_1 - a_0}\right)^{1/\gamma} + (1-w) F(k)
.\end{equation}
Here $a_0$ and $a_1$ are respectively the minimum and maximum input values considered, and $w$ is a user-defined weight set by default to $0.2$ for NAFE and $0.7$ for MGN. In order to have comparable output, we set $\gamma=2.4$ for both NAFE and MGN. The detail layers $F(k)$ of NAFE and MGN depend on additional arbitrary parameters. For NAFE we use the width $N$ of the neighborhood for local histogram equalization and the denoising coefficient $\sigma$ (set to 129 and 12 respectively in the following).  For MGN we use the scaling parameter $k$ of the $\arctan$ function, the synthesis weights (all set to 1 by default), the number of scales (6 by default), and the widths of the Gaussian kernels, which by default follow a geometric progression, similarly to the \atrous transform. Conversely, in WOW, the values of most parameters are dictated by either the use of the \atrous transform (the progression of scales) or the whitening principle (the number of scales, the absence of synthesis weights). The only free parameters are the denoising significance levels. 

As a test case, we picked an HRI$_\text{EUV}$ image recorded on 2022~April~2 at 09:30:55~UT while Solar Orbiter was at 0.38~AU from the Sun. The $\gamma$-stretched layer ($\gamma=2.4$) of NAFE and MGN is shown in the top left panel of Fig.~\ref{fig:wow_nafe_mgn}. An animated version containing 450 frames acquired from 09:19:15 to 10:34:05~UT is available online. For NAFE (middle row) and MGN (bottom row), the output is on the left, and the corresponding detail layer on the right. WOW does not have a $\gamma$-stretched layer, and thus its output (top right panel) must be compared with the detail layers of NAFE and MGN (right column). For all images the gray scale maps to the central 99.8\% of values.

NAFE and MGN produce a somewhat similar output, with NAFE being more contrasted and MGN emphasizing the small scales to a greater degree. The detail layer of MGN presents the typical dark fringes around bright features produced by unsharp masking. Despite the similarities mentioned in Section~\ref{sec:method}, WOW produces a significantly more contrasted output than MGN and no ringing, with notably less noise. Judgment of the merits of the enhancement methods can be subjective when free parameters are involved. The output of WOW is both as sharp as MGN and as contrasted as NAFE, which suggests that the whitening principle may correspond to what is qualitatively considered to be good results.

\subsection{Performance\label{sec:performance}}

\begin{table*}
\caption{Comparison of execution times (in seconds) for NAFE (on 48 cores), MGN, and WOW, for the images presented in this paper.}
\begin{tabular}{lcccccc}
\hline\hline
 & & & & \multicolumn{3}{c}{WOW}\\ \cline{5-7}
Image & size & NAFE & MGN & standard & edge-aware & edge-aware (6 scales) \\
\hline
HRI$_\text{EUV}$ 2022-03-17T04:02:57 & $2048\times2048$ & 98.8 & 7.3 & 2.4 & 11.4 & 7.6 \\
FSI$_\text{304}$ 2022-03-21T06:30:15 & $3040\times3072$ & 936.9 & 11.4 & 10.9 & 37.3 & 16.9 \\
AIA$_\text{174}$ 2012-08-31T19:44:35 & $4096\times4096$ & 813.3 & 49.8 & 13.6 & 53.1 & 27.3 \\
LASCO C2 1998-06-02T13:31:06 & $1024\times1024$ & 190.7 & 1.3 & 0.5 & 2.3 & 1.9 \\
HRI$_\text{EUV}$ 2022-04-02T09:30:55 & $2048\times2048$ & 100.4 & 5.8 & 2.4 & 11.3 & 7.4 \\
FSI$_\text{174}$ 2022-03-17T06:00:45 & $3072\times3072$ & 1330.5 & 13.8 & 10.9 & 37.7 & 17.0 \\
\hline
\end{tabular}
\label{tab:performance}
\end{table*}

WOW was implemented in Python using \verb|open-cv| bindings to compute the convolutions;   the code is available on GitHub\footnote{\url{https://github.com/frederic-auchere/wavelets/}}. We compare the performance to a parallelized NAFE implementation in Python\footnote{\url{https://git.ias.u-psud.fr/ebuchlin/aia-movie/-/blob/master/medocimage/nafe.py}} validated by comparison with the original implementation, and to the \verb|sunkit-image|\footnote{\texttt{sunkit-image} (\url{https://docs.sunpy.org/projects/sunkit-image/}) is a SunPy \citep{sunpy2020} affiliated package.} implementation of MGN. The results of performance tests on a 2.40~GHz Intel Xeon Silver 4214R CPU are summarized in Table~\ref{tab:performance}. The execution time of WOW and MGN depends on the number of scales, which is set to six by default in MGN. In WOW, the number of scales is set to be $\ln(N)/\ln(2)$. For comparison with MGN, we therefore also include execution times for the edge-aware version of WOW limited to six scales. The operations required by the two algorithms are very similar, hence the similar performance. However, the bilateral filter  makes the edge-aware version of WOW significantly slower than the standard one. Compared to NAFE, it is much faster in all cases while providing equivalent results.

\subsection{Back to subjectivity}

WOW was developed   to identify an objective criterion for image enhancement that leads to a method free of arbitrary parameters. While the approach seems successful, the whitening of the power spectrum is not necessarily the unique viable choice. In addition, manual adjustment of the output may be desirable. Naturally, a weighted $\gamma$-stretched input image can be added to the WOW image, as in NAFE and MGN. However, any other transfer function can be used to rescale the large dynamic range of the input onto a restricted range of output values. A logarithmic scaling is commonly used for solar EUV images. \cite{Lupton2004} proposed to stretch the data using $\arcsinh(x/\beta)$, which is linear for $x\ll 1$ and logarithmic for $x\gg 1$, with $\beta$ a free parameter. Synthesis weights can also be applied at each scale (Equation~\ref{eq:weighted_synthesis}) to provide enhanced control over the result.

\section{Summary\label{sec:conclusions}}

We introduced a new image enhancement algorithm that produces sharp and contrasted outputs from a variety of images, without arbitrarily determined coefficients nor the addition of a gamma-stretched image. While it is possible to introduce ad hoc parameters at both the decomposition and the synthesis stages, the power spectrum  whitening principle at the heart of the method, although arbitrary in itself, dictates that these parameters be set to one, yielding an optimum parameter-free scheme. The optional denoising step still requires significance levels to be set manually; however, based on the work of~\cite{Batson2019}, it may be possible to determine optimal values automatically. The regular variant is significantly faster than MGN and produces more contrasted results. The sensitivity of the nominal scheme to spikes and edges is mitigated in the edge-aware variant, which is significantly slower, but remains practical on a typical laptop computer.

\begin{acknowledgements}
Solar Orbiter is a space mission of international collaboration between ESA and NASA, operated by ESA. The EUI instrument was built by CSL, IAS, MPS, MSSL/UCL, PMOD/WRC, ROB, LCF/IO with funding from the Belgian Federal Science Policy Office (BELSPO/PRODEX PEA C4000134088); the Centre National d’Etudes Spatiales (CNES); the UK Space Agency (UKSA); the Bundesministerium für Wirtschaft und Energie (BMWi) through the Deutsches Zentrum für Luft- und Raumfahrt (DLR); and the Swiss Space Office (SSO). AIA is an instrument on board SDO, a mission for NASA's Living With a Star program.
SOHO/LASCO data are produced by a consortium of the Naval Research Laboratory (USA), Max-Planck-Institut für Aeronomie (Germany), Laboratoire d'Astronomie (France), and the University of Birmingham (UK). SOHO is a project of international cooperation between ESA and NASA. Gabriel Pelouze's work was supported by a CNES post-doctoral grant.
This work used data provided by the MEDOC data and operations centre (CNES / CNRS / Université Paris-Saclay), \url{http://medoc.ias.u-psud.fr/}.
\end{acknowledgements}

\bibliographystyle{aa}
\bibliography{bibliography}

\end{document}